\documentclass[twocolumn]{aastex63}
\makeatletter
\let\frontmatter@title@above=\relax
\makeatother
\usepackage{lipsum}
\usepackage{graphicx}
\usepackage[normalem]{ulem}
\usepackage[nointegrals]{ wasysym }
\usepackage{amsmath}
\usepackage{tikz}
\usetikzlibrary {shapes.geometric}

\usepackage{hyperref}

\def\cos{\text{cos}}
\def\sin{\text{sin}}
\def\lp{\left(}
\def\rp{\right)}
\def\lb{\left[}
\def\rb{\right]}
\def\t{\text}

\accepted{\today}
\shorttitle{Super-Chandrasekhar Supernovae}
\shortauthors{Fitz Axen et al.}

\graphicspath{{./}{figures/}}

\begin{document}

\title{The Progenitors of Superluminous Type Ia Supernovae}

\correspondingauthor{Margot Fitz Axen}
\email{fitza003@utexas.edu}

\author[0000-0001-7220-5193]{Margot Fitz Axen}
\affiliation{Department of Astronomy, The University of Texas at Austin, 2515 Speedway, Stop C1400, Austin, Texas 78712-1205, USA.}

\author[0000-0002-3389-0586]{Peter Nugent}
\affiliation{
Lawrence Berkeley National Laboratory, Berkeley, California 94720, USA.
Department of Astronomy, University of California, Berkeley, CA, 94720, USA.}

\begin{abstract}
Recent observations of type Ia supernovae (SNe Ia) have discovered a subclass of `super-Chandrasekhar’ SNe Ia (SC SNe Ia) whose high luminosities and low ejecta velocities suggest that they originate from the explosions of white dwarfs (WDs) with masses that exceed the Chandrasekhar mass limit. Different models have been proposed to explain the progenitors of these explosions, including a ‘magnetized WD’ model and a ‘WD merger’ model. To test the robustness of these models, we conduct a 1D numerical parameter survey of WD explosions using these models as initial conditions. We follow the explosions using the hydrodynamics code Castro and then use the radiation transport code SuperNu to create light curves and spectra for the models. We find that while both classes of models fall within the range of SC SNe Ia observations on the light curve width-luminosity relation, only the WD merger models reproduce the observed low ejecta velocities. The light curves of our merger models are more similar photometrically  to observations than our magnetized models. Given this, we discuss possible explanations for the brightest SC SNe Ia observations that cannot be reproduced with our WD merger models. This study provides the basis for future SC SNe Ia observations and higher-dimensional numerical models. 
\end{abstract}



\section{Introduction}

Type Ia supernovae (SNe~Ia) are used as cosmological ‘standard candles’ due to the homogenity of their light curves, which follow a characteristic relationship between their peak luminosity and width \citep{phillips_1993}. Observations of high-redshift SNe~Ia prooved the accelerating expansion of the universe and  were used to make the first measurements of the cosmological constant \citep{perlmutter_1999, riess_observational_1998}. However, in recent decades, many subtypes of SNe~Ia have been discovered that do not follow the standard width-luminosity relation and must be excluded from cosmological surveys. These include 1991bg-like SNe \citep{filippenko_1992_1991bg}, 2002cx-like SNe \citep{li_2003}, 1991T-like SNe \citep{filippenko_1992_199T}, and 2003fg-like SNe \citep{ howell_type_2006, hicken_luminous_2007, yamanaka_early_2009, scalzo_nearby_2010, yuan_exceptionally_2010, silverman_fourteen_2011,  taubenberger_extremes_2017, taubenberger_sn_2019, chen_2019_ASASSN_15pz, ashall_carnegie_2021, jiang_discovery_2021, dimitriadis_2022}. These SNe subtypes differ spectroscopically and photometrically from Branch-normal SNe~Ia \citep{branch_2006}. 

One of the rarest subtypes is  2003fg-like SNe, which are often dubbed ``super-Chandrasekhar" SNe~Ia (hereafter SC SNe Ia). They have slowly declining light curves ($\Delta m_{15} (B) < 1.3$ mag) and are exceptionally luminous, with peak absolute B-band magnitudes of $-19 < M_B < -21$ mag \citep{ashall_carnegie_2021}. Spectroscopically, they exhibit strong CII lines a few days after explosion and weaker SiII and FeIII lines than normal SNe~Ia \citep{taubenberger_high_2011, dimitriadis_2022}. Their linewidths suggest unusually low ejecta velocities \citep{howell_type_2006}. Modeling these SNe analogously to normal SNe suggests that many of these SNe originiate from WDs that exceed the Chandrasekhar mass ($M_{ch}$). 

 All SNe~Ia originate from the explosion of an unstable carbon-oxygen (CO) white dwarf (WD) accreting mass in a binary system \citep{nugent_supernova_2011}, but the exact nature of their progenitor remains unknown. They may exist in a `single degenerate' (SD) system in which their companion star is a normal star, or a `double degenerate' (DD) system in which their companion star is another WD \citep{hillebrandt_towards_2013}. Both of these progenitor classes may be able to form super-Chandrasekhar mass systems that would explain the 2003fg-like observations. 
 
 In the single-degenerate case, the WD properties must enable it to remain stable at masses greater than $M_{ch}$. This may be possible if the WD is highly magnetized, rotating, or both, as we describe below. Observations from the Sloan Digital Sky Survey (SDSS) shows that $\approx$ 10 \% of WDs are magnetized, with surface magnetic fields in the range $10^4-10^9$ G, \citep{schmidt_2003_sdss} and suggests that their masses are higher than their non-magnetized counterparts \citep{vanlandingham_2005_sdss}. The internal magnetic fields of WDs are not known, but they are expected to be higher than their surface magnetic fields, and may reach $\approx$ $10^{14}$ G in their center \citep{franzon_effects_2015, chatterjee_maximum_2017, otoniel_strongly_2019}. With these magnetic fields, the WD may reach a maximum mass of $\approx$ 2.0 solar masses or greater \citep{franzon_effects_2015, otoniel_strongly_2019, das_new_2013, das_maximum_2014, bhattacharya_effect_2021, bhattacharya_2022_magnetic}. Additionally, rotation due to rapid accretion can provide a similar opportunity for WDs to remain stable at masses greater than $M_{ch}$ \citep{yoon_evolution_2005, franzon_effects_2015}. While these WDs have been proposed as potential SC SNe Ia progenitors, they have not been studied in numerical SNe~Ia simulations.
 
 In the double degenerate case, a system of merging WDs that has a total mass greater than $M_{ch}$ may also produce SC SNe Ia. The system may explode violently soon after contact \citep{pakmor_sub-luminous_2010, pakmor_violent_2011, pakmor_normal_2012, pakmor_helium-ignited_2013} or quiescently after the secondary has been completely disrupted \citep{dan_prelude_2011,raskin_2013, raskin_2014, noebauer_2016_csm}. Studies have shown that these WD merger systems may be the progenitors of normal and subluminous SNe~Ia; however, they have not been shown to be potential progenitors of superluminous SNe~Ia to date. 
 
 This paper presents a survey of super-Chandrasekhar mass explosion models and explores their observational relevance to 2003fg-like SNe. We explore models of both highly-magnetized WD models and WD merger models using calculations and simulation results from previous papers. In Section \ref{section:methods} we describe our WD models and numerical methods. In Section \ref{section:results} we describe the results of our simulations, focusing on the differences between the different models in both the hydrodynamic properties of the explosions and the observable spectroscopic and photometric output.  
 We discuss our results and summarize our conclusions in Section \ref{section:discussion}.

\section{Numerical Methods}
\label{section:methods}

We model the explosion in several stages. First, the initial model is built by starting with a WD progenitor model of either a WD merger system or highly magnetized WD. After the initial model is constructed, a detonation is initiated in the center of the star. The resulting thermonuclear hydrodynamics are then evolved in 1D. Each model is evolved until the ejecta reaches homologous expansion, after which we perform radiative transport calculations to produce light curves and spectra for each model.

\subsection{Hydrodynamics and Nuclear Processes}

The compressible Eulerian hydrodynamics code Castro is used to follow the progression of the explosions \citep{almgren_castro_2010, zingale_meeting_2018}. We enable Castro's built in monopole gravity and the Helmholtz equation of state \citep{timmes_2000_helmholtz}. A 13 isotope $\alpha$-chain nuclear network is used to monitor the nuclear reactions \citep{timmes_1999_network}. The network includes $(\alpha, \gamma)$ and $(\gamma, \alpha)$ reactions for 13 isotopes: $^4$He, $^{12}$C, $^{16}$O, $^{20}$Ne, $^{24}$Mg, $^{28}$Si, $^{32}$S, $^{36}$Ar, $^{40}$Ca, $^{44}$Ti, $^{48}$Cr, $^{52}$Fe, and $^{56}$Ni. 

Additionally, we perform a calculation with a 21 isotope reaction network which includes all of the isotopes in the 13 reaction network as well as $^1$H, $^3$He, $^{14}$N, $^{56}$Cr, $^{54}$Fe, and $^{56}$Fe. For this test, we use one of the lowest density merger models, MG051018, as the lower density models produce less heavy elements and are more likely to be impacted by the choice of reaction network. In this example, the larger network did not yield significantly different results and it was determined that the 13 isotope network was sufficient for the scope of this study. 

Castro allows for adaptive mesh refinement (AMR), which enables the increase of resolution in areas of interest in the simulation.  However, we find that performance in the code is best on GPUS if running without AMR and instead starting with a high number of grid cells. For all of our models, we use ~ $10^5$ grid cells over a range of $10^5$ km, or about 1 km/cell. We perform a resolution test to verify that this resolution is sufficient using the MAG1010 magnetized model, and found that increasing the number of grid points did not impact the results of the calculation. 

\subsection{WD Models}

The initial models were created using a different process for the WD mergers and the magnetized WDs. For the WD merger models, we construct an isothermal WD as the primary accretor and then add mass around it to model the accretion of the secondary WD. For the magnetized WD models, we construct a single WD with a modified EOS to  account for the magnetic field. For both models, the composition  of the WD(s) is 50 \% carbon and 50 \% oxygen.

\subsubsection{Initial Models: WD Merger Models}

To construct a 1D model for a WD merger we follow a two step process to compute the stucture of the primary star and then the secondary star being accreted around it. We choose as input the central density $\rho_c$ of the primary star and integrate the equations of stellar structure from the center to the surface of the star, which we take to be $\rho_s=10^{-4} \rm g/cm^3$.  We use a semi-relativistic equation of state which interpolates between the non-reletevistic pressure $P_{\t{deg}, nr}$ and reletevistic pressure $P_{\t{deg}, r}$ as \citep{paczynski_1983_pressure}
\begin{equation}
    P_{\rm deg}^{-2} = P_{\t{deg}, nr}^{-2}+ P_{\t{deg}, r}^{-2}.
\end{equation}
Here the two limits are $P_{\t{deg}, nr} = K_{ nr}\rho^{4/3}$ and $P_{\t{deg}, r} = K_{r}\rho^{5/3}$ where $K_{ nr}$ and $K_{r}$ are polytropic constants in the non-relativistic and relativistic limits respectively.

For the structure of the star, we integrate the Tolman-Oppenheimer-Volkoff equation \citep{oppenheimer_1939}:
\begin{equation}
    \frac{d\rho}{dr}=-\frac{G[\rho+P/c^2][m(r)+4\pi r^3P/c^2]}{[r^2-2Gm(r)r/c^2](dP/d\rho)},
    \label{eq:tolman}
\end{equation}
and the equation of hydrostatic equilibrium:
\begin{equation}
    \frac{dm}{dr}=4\pi r^2\rho.
    \label{eq:hydrostatic}
\end{equation}
This integration gives a relationship between the mass and the radius of the primary that is approximately equal to that derived by Chandrasekhar \citep{bhattacharya_2022_magnetic}:
\begin{equation}
    R_P \approx 9 \times 10^3 \lp \frac{M_P}{M_{\astrosun}}\rp ^{-1/3}\sqrt{1-\lp \frac{M_P}{M_{ch}}\rp^{4/3}} \rm km,
    \label{eq:mass_radius}
\end{equation}
although with negligible differences due to our choice of $\rho_s=10^{-4} \rm g/cm^3$.

Once we have constructed the models for the primaries, we construct the model for the secondary WD by assuming that the merging disrupts the outer layers of the primary and creates an outer medium at approximately constant density $\rho_{\rm add}$.  We remove the outer profile of the primary where $\rho<\rho_{\rm add}$ and instead add mass onto the primary at $\rho=\rho_{\rm add}$ until the total mass of the configuration is equal to a total mass $M_{\rm tot}=M_P+M_{SD}$, where $M_{SD}$ is the mass of the secondary star. We keep a constant pressure equal to the pressure of the primary at $\rho=\rho_{\rm add}$ throughout $M_{SD}$. Note that because $\rho_{\rm add}$ is orders of magnitude lower than $\rho_c$, this process removes a negligible amount of mass ($< 1 \% $) from the primary. The resulting mass-radius profile for one of our models is shown in Figure \ref{fig:mass_radius_profiles} by the solid blue line.

This parameter survey varies the central density of the primary $\rho_c$ (effectively varying the primary mass), the total mass of the configuration $M_{\rm tot}$ and the density of the added secondary material $\rho_{\rm add}$. Varying the density of the accreted material is intended to explore the difference between very concentrated accreted material vs more diffuse.  We only consider models in which $M_P$ and $M_{SD}$ are both in the range of 0.8-1.2 $M_{\astrosun}$, at which both stars are expected to be CO WDs \citep{dan_prelude_2011}. Additionally, we have the constraints $M_{ SD}<M_P$ and $M_{\rm tot} > M_{\rm ch}$. Our full parameter space of merger models is shown in Table \ref{table:simulation_models_mergers}.

\begin{table*} 
\small
\centering
\begin{tabular}{|c c c c c c c c c c|}
\hline
Run Name & Symbol & $\rm \rho_c$ [$\rm g/cm^3$] & $\rho_{\rm add}$ [$\rm g/cm^3$] & $M_{\rm tot}$ [$M_{\astrosun}$] & $M_P$ [$M_{\astrosun}$] & $R_P$ [km] & $M_{SD}$ [$M_{\astrosun}$] & $q$ & $R_{\rm tot}$ [km] \\
\hline
\hline
MG050518 & \includegraphics[width=0.15in]{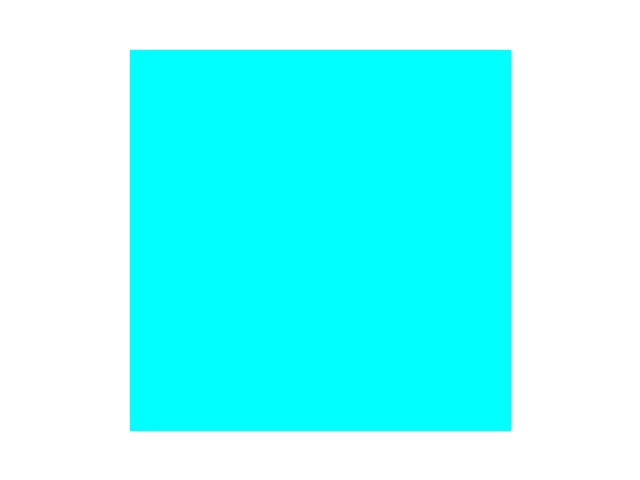} & $5 \times 10^7$ & $5 \times 10^4$ & 1.8 & 1.09 & 5065 & 0.71 & 0.65 & 18963  \\
\hline
MG051018 & \includegraphics[width=0.15in]{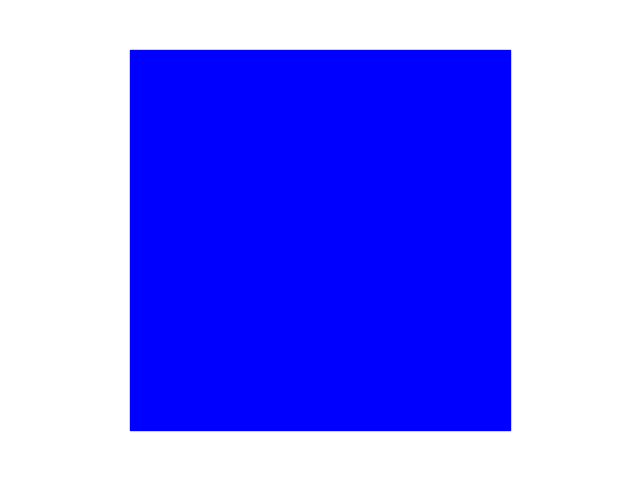} & $5 \times 10^7$ & $10^5$ & 1.8 & 1.09 & 5065 & 0.71 & 0.65 & 15133  \\
\hline
MG055018 & \includegraphics[width=0.15in]{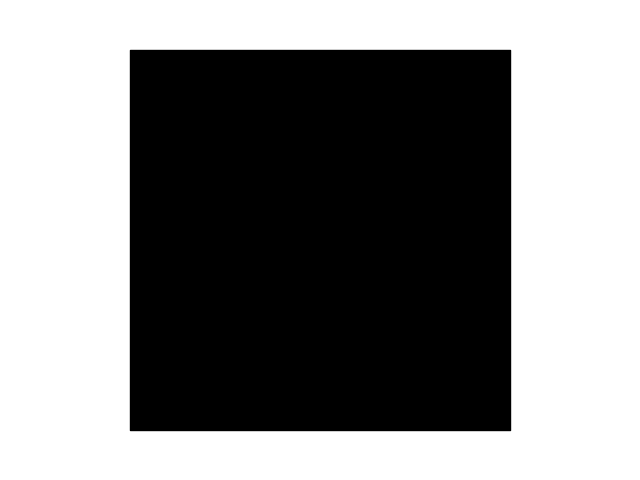} & $5 \times 10^7$ & $5 \times 10^5$ & 1.8 & 1.09 & 5065 & 0.71 & 0.65 & 9161  \\
\hline
MG050520 & \includegraphics[width=0.15in]{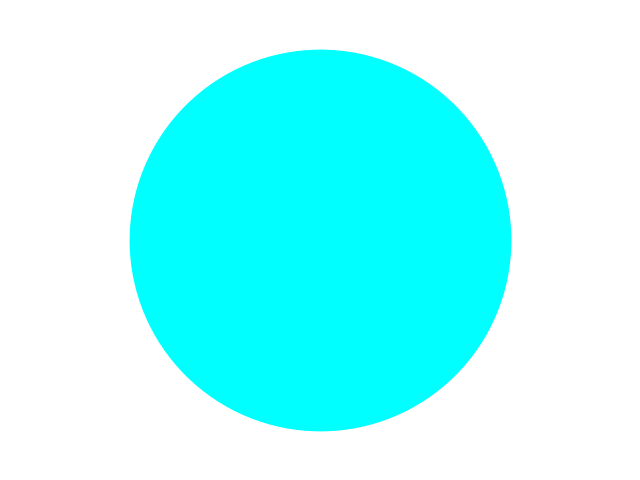} & $5 \times 10^7$ & $5 \times 10^4$ & 2.0 & 1.09 & 5065 & 0.91 & 0.83 & 20581 \\
 \hline
MG051020 & \includegraphics[width=0.15in]{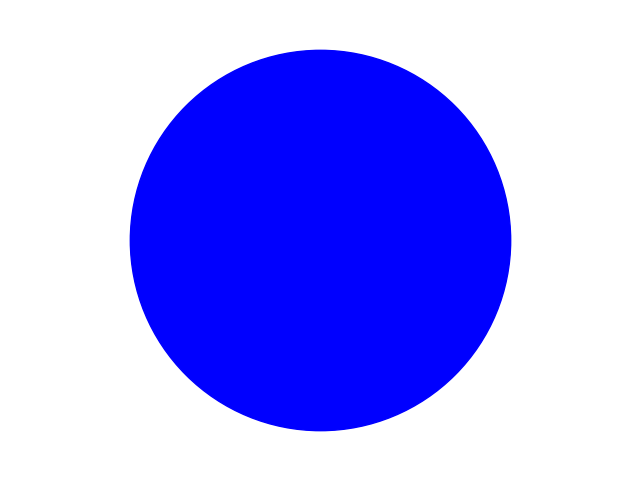} & $5 \times 10^7$ & $10^5$ & 2.0 & 1.09 & 5065 & 0.91 & 0.83 & 16406  \\
\hline
MG055020 & \includegraphics[width=0.15in]{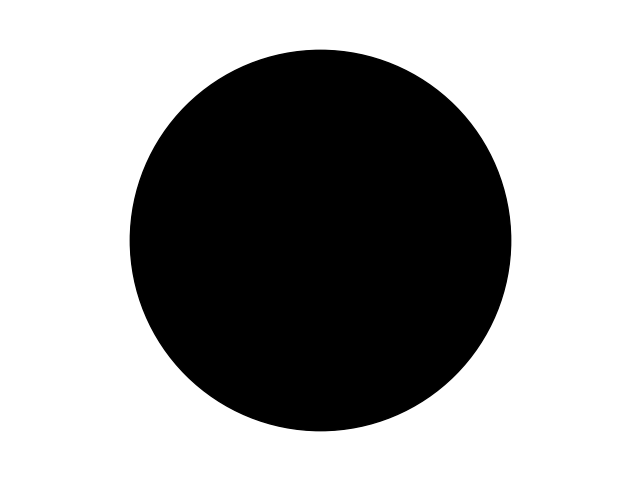} & $5 \times 10^7$ & $5 \times 10^5$ & 2.0 & 1.09 & 5065 & 0.91 & 0.83 & 9861  \\
  \hline
  MG100518 & \includegraphics[width=0.15in]{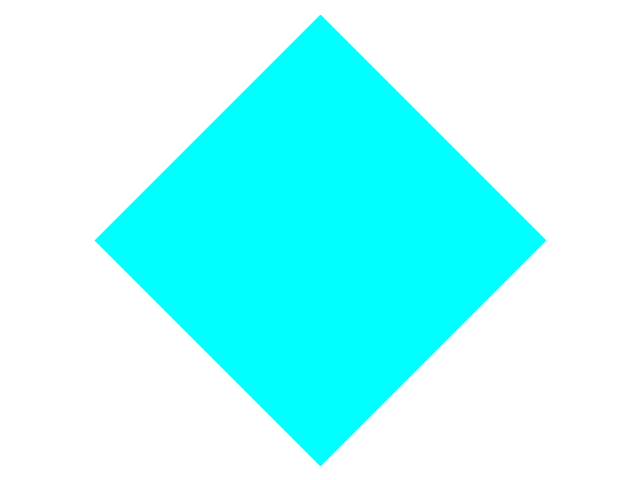} & $1 \times 10^8$ & $5 \times 10^4$ & 1.8 & 1.18 & 4341 & 0.62 & 0.53 & 18102 \\
 \hline
MG101018 & \includegraphics[width=0.15in]{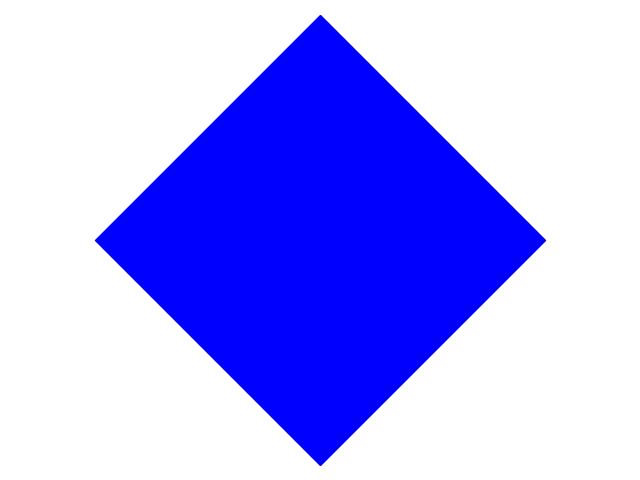} & $1 \times 10^8$ & $10^5$ & 1.8 & 1.18 & 4341 & 0.62 & 0.53 & 14426  \\
\hline
  MG105018 & \includegraphics[width=0.15in]{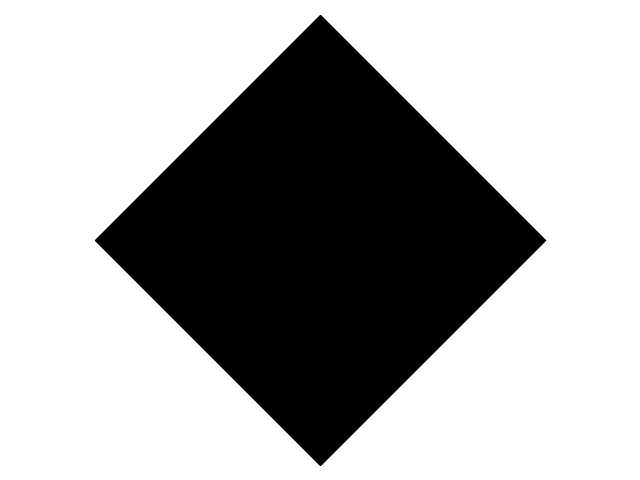} & $1 \times 10^8$ & $5 \times 10^5$ & 1.8 & 1.18 & 4341 & 0.62 & 0.53 & 8668  \\
 \hline
 MG100520 & \includegraphics[width=0.15in]{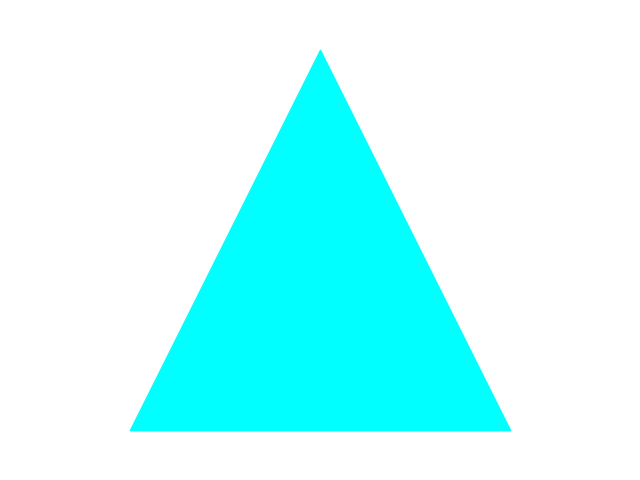} & $1 \times 10^8$ & $5 \times 10^4$ & 2.0 & 1.18 & 4341 & 0.82 & 0.69 & 19859 \\
 \hline
MG101020 & \includegraphics[width=0.15in]{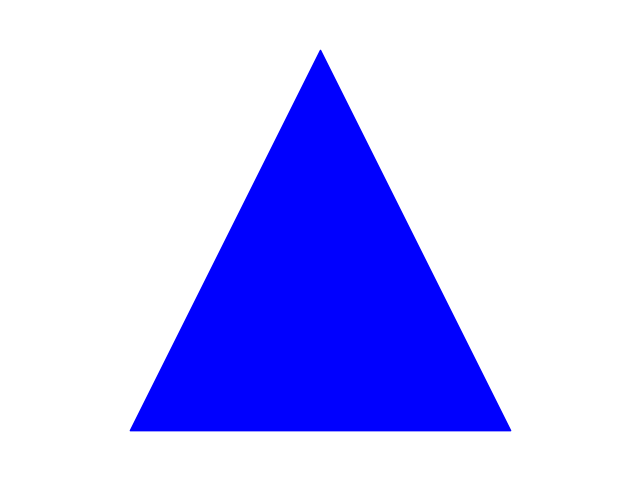} & $1 \times 10^8$ & $10^5$ & 2.0 & 1.18 & 4341 & 0.82 & 0.69 & 15811  \\
 \hline
MG105020 & \includegraphics[width=0.15in]{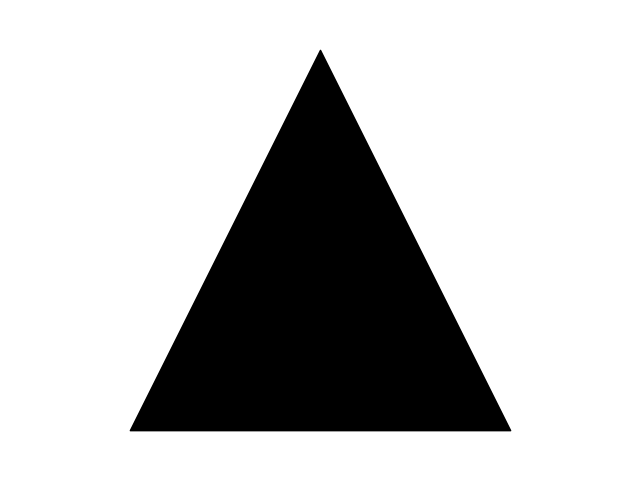} & $1 \times 10^8$ & $5 \times 10^5$ & 2.0 & 1.18 & 4341 & 0.82 & 0.69 & 9440  \\
\hline
MG100522 & \includegraphics[width=0.15in]{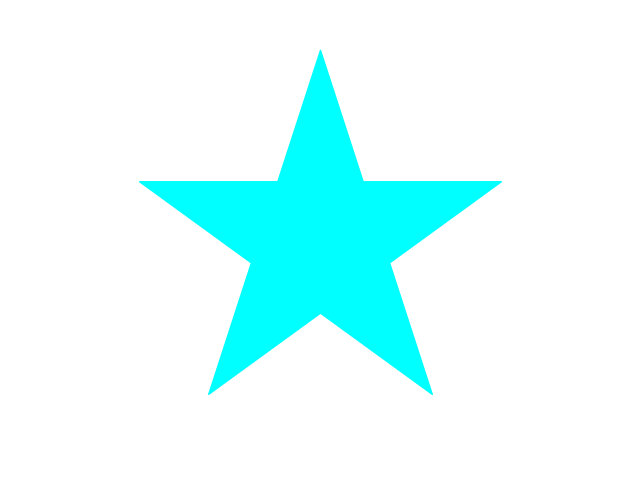} & $1 \times 10^8$ & $5 \times 10^4$ & 2.2 & 1.18 & 4341 & 1.02 & 0.86 & 21349 \\
\hline
MG101022 & \includegraphics[width=0.15in]{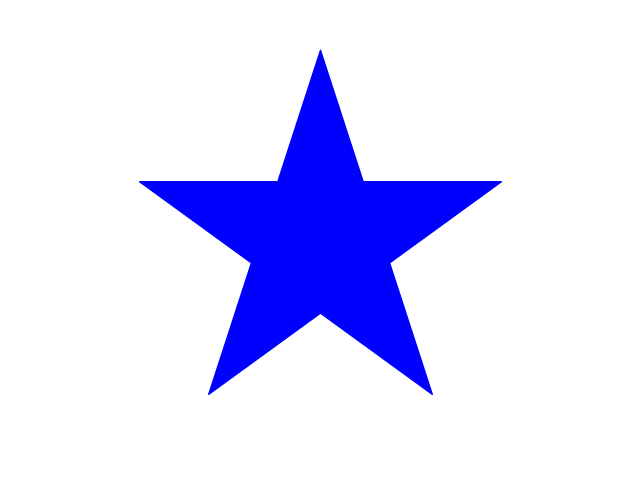} & $1 \times 10^8$ & $10^5$ & 2.2 & 1.18 & 4341 & 1.02 & 0.86 & 16987  \\
\hline
MG105022 & \includegraphics[width=0.15in]{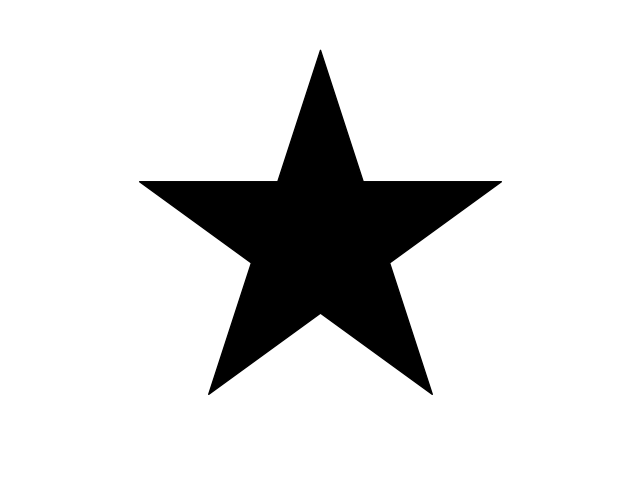} & $1 \times 10^8$ & $5 \times 10^5$ & 2.2 & 1.18 & 4341 & 1.02 & 0.86 & 10102  \\
  \hline
 \end{tabular}
\caption{Summary of the merger models run. We denote the runs `MGxxyyzz', where xx represents the central density in units of $10^8 \rm g/cm^3$, yy represents the secondary material density in units of $10^5 \rm g/cm^3$, and zz represents the total mass of the configuration. Columns are the simulation name, plot symbol, central density, density of the added material, total mass of the configuration,  mass of the primary, radius of the primary, mass of the secondary, mass ratio, and total radius. The symbol used for each simulation  distinguishes the different combinations of [$\rho_c$,$M_{\rm tot}$] (columns 3 and 5) by shape and the value of $\rho_{\rm add}$ (column 4) by color.}
\label{table:simulation_models_mergers}
\end{table*}

\subsubsection{Initial Models: Highly Magnetized WDs}

To construct the models for the magnetized WDs, we follow the model in \cite{bhattacharya_2022_magnetic}. The addition of a magnetic field modifies the EOS of the star. We use the magnetic field model used in \cite{bhattacharya_2022_magnetic} which has been used extensively to model magnetized neutron stars and WDs \citep{das_maximum_2014}:
\begin{equation}
    B\lp\frac{\rho}{\rho_0}\rp=B_s+B_0\lb1-\rm exp\lp-\eta \lp\frac{\rho}{\rho_0}\rp^{\gamma}\rp\rb.
    \label{eq:magnetic_field}
\end{equation}
Here $B_s$ is the surface magnetic field, $B_0$ is a fiducial magnetic field, and $\eta$ and $\gamma$ are dimensionless parameters that determine how the magnetic field changes from the core to the surface. For our calculations here, we set $\rho_0=10^9 \rm g cm^{-3}$, $\eta=0.8$, and $\gamma=0.9$ for all calculations, following \cite{bhattacharya_2022_magnetic}. We set the surface magnetic field to $B_s=10^{7}$ G as this parameter has negligible effect on the profile at high WD masses \citep{bhattacharya_2022_magnetic}. The profile in Equation \ref{eq:magnetic_field} indicates the magnitude of the magnetic field at various density points throughout the star and hence radial coordinates.

\begin{figure}[th!]
\centering
\includegraphics[width=0.49 \textwidth]{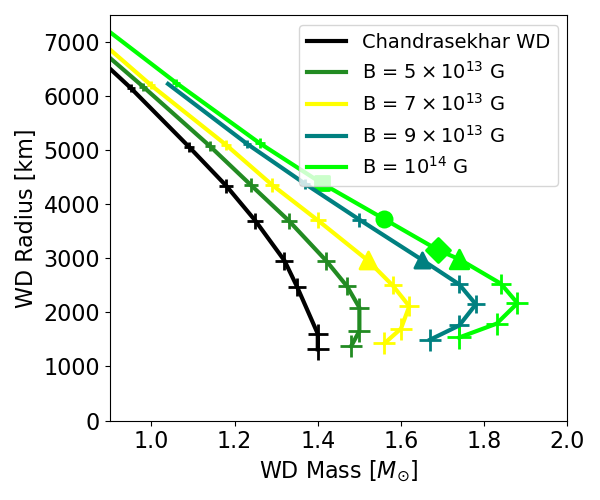}
\caption{The mass radius relation for several different values of the WD central magnetic field (represented by the different colored lines). The crosses show different values of central density $\rho_c$ between $10^8 \rm g/cm^3$ and $10^{10} \rm g/cm^3$, with higher values being further to the right on each line. All other symbols are models that we test, listed in Table \ref{table:simulation_models_magnetic}.}
\label{fig:magnetic_models}
\end{figure}

\begin{figure}[th!]
\centering
\includegraphics[width=0.49 \textwidth]{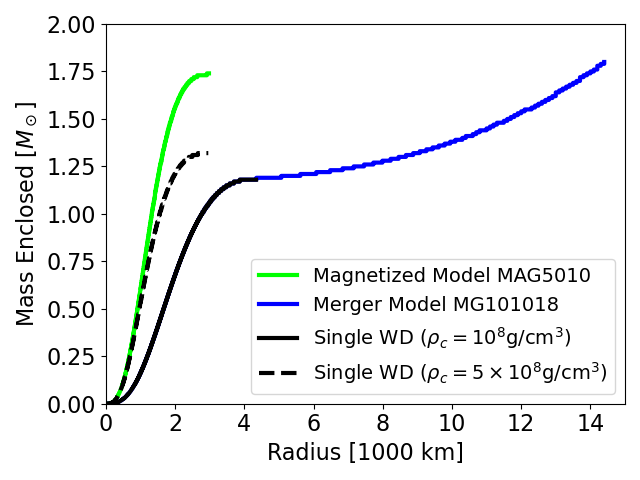}
\caption{The mass radius profiles for two single WDs, our MG101018 merger model, and our MAG5010 magnetized model.The merger model has the same profile at inner radii as the single WD of the same central density, while the magnetized model has a modified EOS that changes its profile throughout.}
\label{fig:mass_radius_profiles}
\end{figure}

Similarly to the method we use to construct the primaries for the WD merger models, we integrate outward from the center of the star at $\rho=\rho_c$ until $\rho=\rho_s=10^{-4} \rm g cm^{-3}$. We integrate Equations \ref{eq:tolman} and \ref{eq:hydrostatic} with a modified EOS where $P = P_{\rm deg}+P_B$ and $\rho = \rho_{\rm mat}+\rho_B$. Here $P_B=B^2/(8 \pi)$ is the magnetic pressure and $\rho_B = B^2/(8\pi c^2)$ is the magnetic density at the appropriate position in the star, with the magnetic field determined using Equation \ref{eq:magnetic_field}. Figure \ref{fig:magnetic_models} shows the resulting mass-radius relation for several different values of $B_0$ for comparison to the Chandrasekhar result. Figure \ref{fig:mass_radius_profiles} shows the resulting mass-radius profile for one of our magnetized models (green line) along with an unmagnetized WD of the same central density for comparison. 

This parameter survey varies the central density $\rho_c$ and the central magnetic field $B_0$. We explore several values for the central density using $B_0= 10^{14}$ G. At higher values for the central magnetic field, the WD is expected to be non-spherical, making it insufficient to be tested with our 1D models \citep{bhattacharya_2022_magnetic}. We also test a couple models at a fixed central density for lower values of $B_0$, at which the WD EOS is still significantly different than for the Chandrasekhar result. Our only constraint for our models is that $M_*>M_{ch}$. Our full parameter space of magnetic models tested is shown in Table~\ref{table:simulation_models_magnetic}. The stars in Figure \ref{fig:magnetic_models} are models that we test.

\begin{table*} 
\small
\centering
\begin{tabular}{|c c c c c c|}
\hline
Run Name & Symbol & $B_0$ [G] & $\rm \rho_c$ [$\rm g/cm^3$] & $M_*$ [$M_{\astrosun}$]  & $R_*$ [km] \\
\hline
MAG1010 & \includegraphics[width=0.15in]{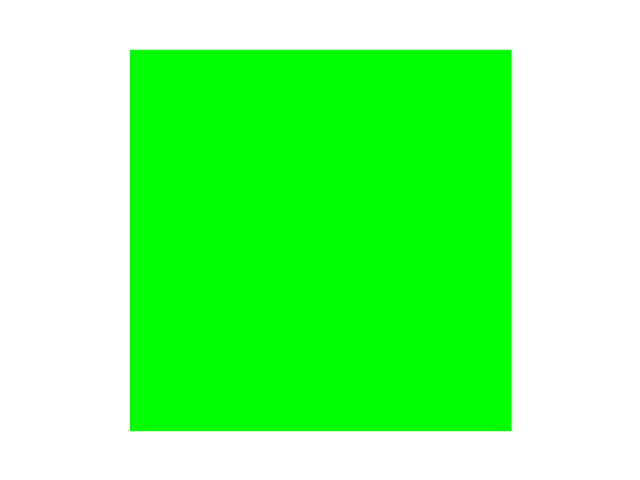} & $10^{14}$ & $1 \times 10^8$ & 1.41 & 4386 \\
\hline
MAG2010 & \includegraphics[width=0.15in]{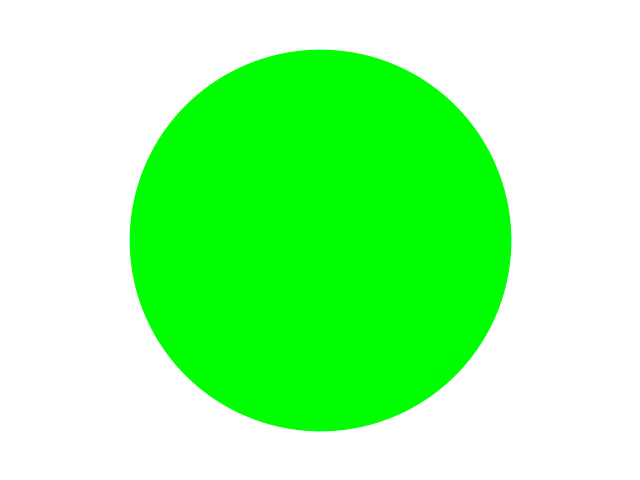} & $10^{14}$ & $2 \times 10^8$ & 1.56 & 3724 \\
\hline
MAG4010 & \includegraphics[width=0.15in]{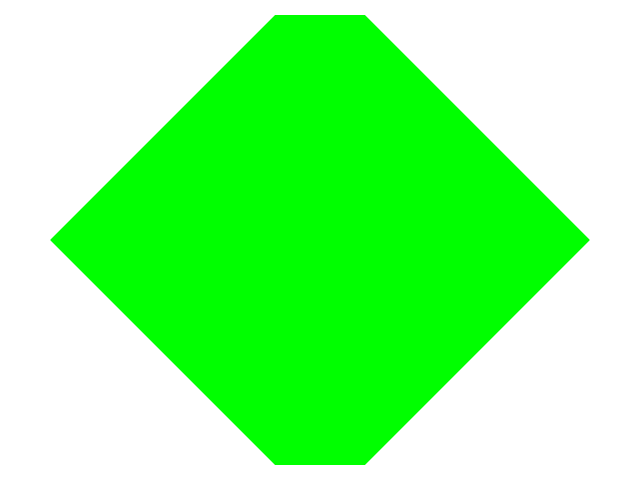} & $10^{14}$ & $4 \times 10^8$ & 1.69 & 3150  \\
\hline
MAG5010 & \includegraphics[width=0.15in]{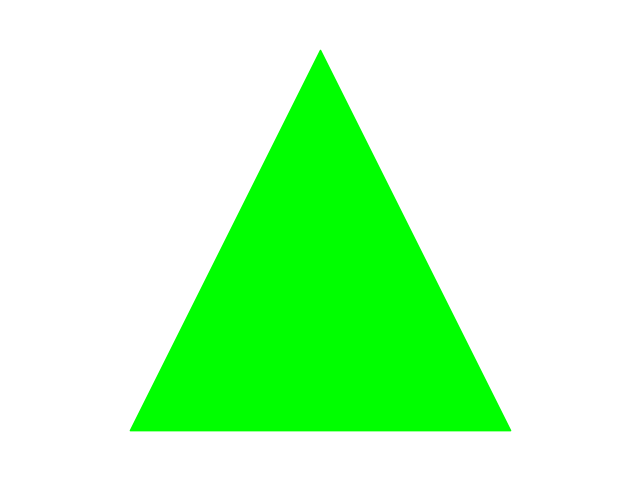} & $10^{14}$ & $5 \times 10^8$ & 1.74 & 2644  \\
\hline
MAG5007 & \includegraphics[width=0.15in]{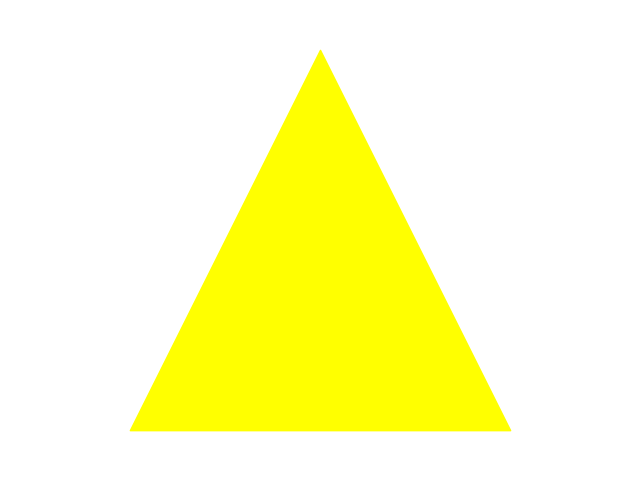} & $7 \times 10^{13}$ & $5 \times 10^8$ & 1.52 & 2598  \\
\hline
MAG5009 & \includegraphics[width=0.15in]{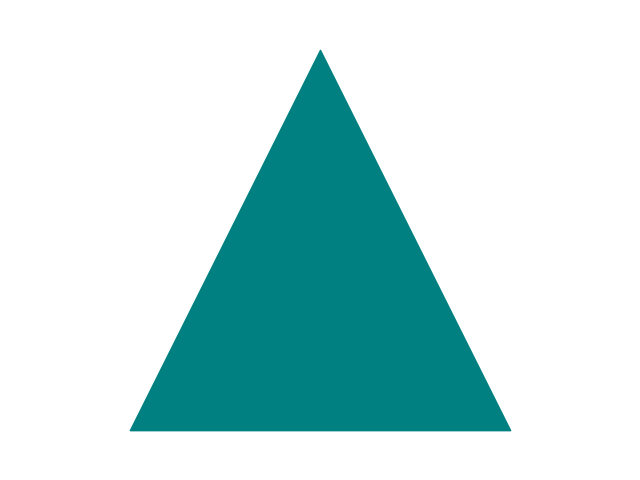} & $9 \times 10^{13}$ & $5 \times 10^8$ & 1.65 & 2626  \\
\hline
 \end{tabular}
\caption{Summary of the magnetized models run. We denote the runs `MAGxxyy', where xx represents the central density in units of $10^8 \rm g/cm^3$ and yy represents the central magnetic field in units of $10^{14}$ G. Columns are the simulation name, plot symbol, central magnetic field, central density, mass of the star, and radius of the star. The symbol used for each simulation distinguishes the value of $B_0$ (third column) by color and the value of $\rho_c$ (fourth column) by shape.}
\label{table:simulation_models_magnetic}
\end{table*}

\subsection{Hydrodynamics and Radiation Transport}

After an initial model is constructed it is imported into Castro where the hydrodynamics calculations are performed. An ignition is triggered in the center of the star by heating a grid cell. Once nucleosynthesis is initiated, all artificial heating is turned off and the hydrodynamics are evolved through homology.

After the SN ejecta reaches homologous expansion we use the SuperNu code \citep{wollaeger_2013_supernu1, wollaeger_2014_supernu2} to create synthetic light curves and spectra for each model. SuperNu is a multi-dimensional time dependant radiation transport code that uses Monte Carlo methods to propagate photons. The calculations are performed under the assumption of local thermodynamic equilibrium (LTE) to determine ionization and excitation fractions in the ejecta. Energy is generated through the radioactive decay chain: $^{56}$Ni $\rightarrow^{56}$Co $\rightarrow^{56}$Fe. 

\section{Results}
\label{section:results}

\subsection{Nucleosynthetic Yields and Kinetic Energy}

\begin{figure*}
\centering
\includegraphics[width=0.49 \linewidth]{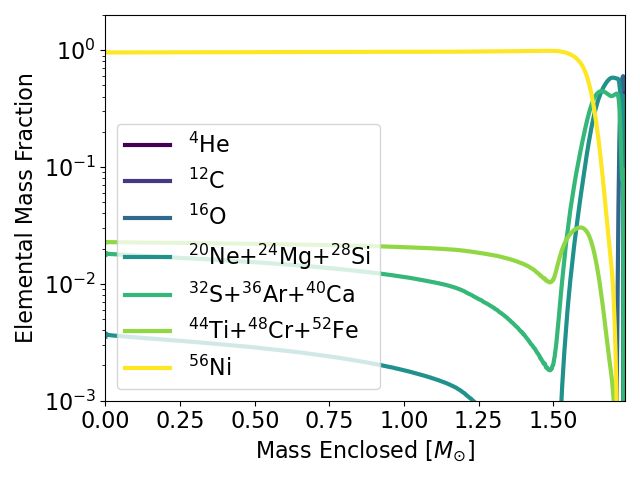}
\includegraphics[width=0.49 \linewidth]{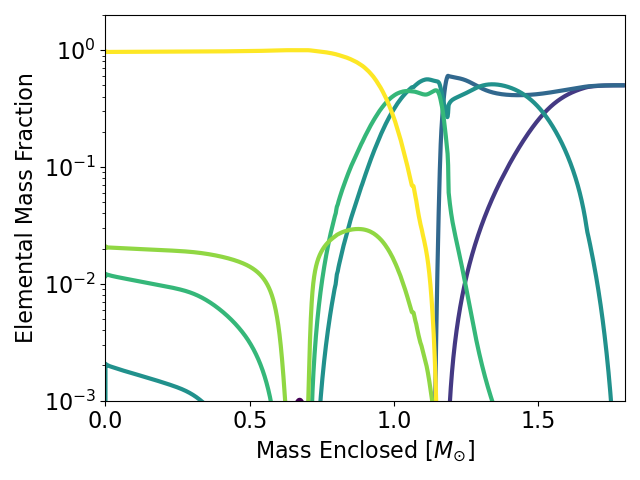}
\caption{Elemental composition of the supernova ejecta as a function of interior mass for the MAG5010 magnetized WD model (left) and the MG101018 WD merger model (right). Even though the merger model has a larger total mass, it doesn't produce as much radioactive Nickel and has a large amount of unburned Carbon and Oxygen. }
\label{fig:mass_fractions}
\end{figure*}

\begin{figure}[th!]
\centering
\includegraphics[width=0.49 \textwidth]{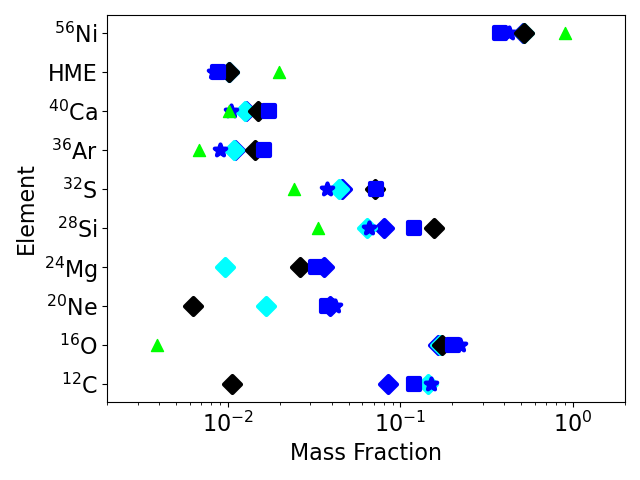}
\caption{Elemental composition of five of our merger models (blue/black) and one of our magnetized models (green), with symbols shown in Tables \ref{table:simulation_models_mergers} and \ref{table:simulation_models_magnetic}. The second row from top shows the three higher mass elements $^{44}\rm Ti$, $^{48}\rm Cr$, and $^{52}\rm Fe$ as one point for each of the models. Missing points exist where the elemental mass fraction for a model is less than $10^{-3}$. }
\label{fig:mass_fractions_scatter}
\end{figure}

Figure \ref{fig:mass_fractions} shows the ejecta composition for the MG101018 merger model (right) and the MAG5010 magnetized model (left) shown in Figure \ref{fig:mass_radius_profiles} when the burning has reached homologous expansion to highlight the general properties of our models. The magnetized model is composed almost entirely of $^{56}\rm Ni$ (yellow line), except at the outer parts of the star. In contrast, though it has a higher total mass than the magnetized model, the merger model has a lower central density and only burns part of the primary to $^{56}\rm Ni$. The secondary material is partially burned to intermediate mass elements, but much of it remains as unburned $^{12} \rm C$ and $^{16}\rm O$. Figure \ref{fig:mass_fractions_scatter} shows the elemental composition of these two models by the royal blue diamonds (merger model) and lime-green triangles (magnetized model). The magnetized model has a higher fraction of $^{56}\rm Ni$ and the higher mass elements $^{44}\rm Ti$, $^{48}\rm Cr$, and $^{52}\rm Fe$, while the merger model has a higher fraction of all of the lower mass elements. 

A natural conclusion of the  secondary material of the merger model burning incompletely is that the mass and density configuration of the secondary have no impact on the amount of $^{56} \rm Ni$ or higher mass elements. Therefore, all of our merger models with the same primary produce the same amount of $^{56} \rm Ni$. The MG05yyzz models make 0.68 $M_{\astrosun}$, while the MG10yyzz models make 0.92 $M_{\astrosun}$ of $^{56} \rm Ni$. This is further highlighted in Figure \ref{fig:mass_fractions_scatter}, which also shows the MG105018, MG100518, MG101022, and MG051018 models. The merger models have different amounts of intermediate mass elements, but the four merger models with the same primary have the same amount of $^{56}\rm Ni$ and high mass elements. Therefore, the three with the same secondary mass (models MG10yy18) have the same mass fraction of $^{56}\rm Ni$. 

\begin{figure}[th!]
\centering
\includegraphics[width=0.49 \textwidth]{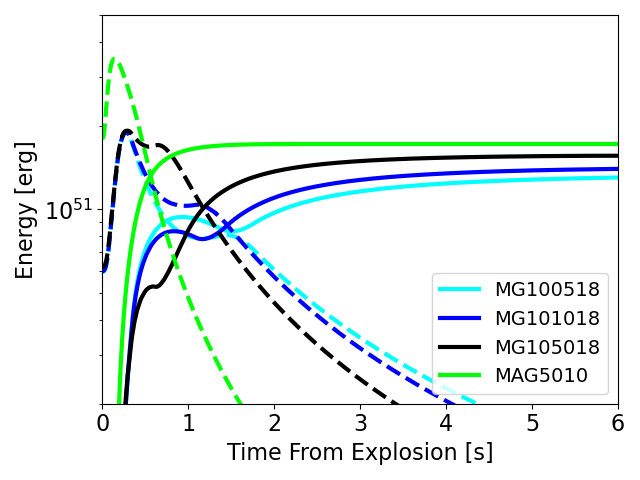}
\caption{Kinetic (solid) and internal (dashed) energies of the supernova ejecta as a function of time for the MG100518, MG101018, and MG105018 WD merger models and MAG5010 magnetized WD model.}
\label{fig:kinetic_energy}
\end{figure}

Figure \ref{fig:kinetic_energy}  shows the kinetic energy as a function of time for the MG10yy18 merger models and the MAG5010 magnetized model. For all models, the total energy is initially dominated by the binding energy of the WD, which is converted to kinetic energy as the WD ejecta is blown outward. However, the magnetized model kinetic energy increases smoothly, while the merger models are stalled by the dense material on the outside. As a result, the merger models reach lower kinetic energies than the magnetized model (even though they have a higher total mass) and take longer to reach homologous expansion.

The kinetic energy profiles also differ between the different density configurations for the merger models. There is initially an inverse correlation between the density of added material and the kinetic energy of the ejecta, because a higher density secondary does not allow the ejecta to expand as quickly. However, as shown in Figure \ref{fig:mass_fractions_scatter}, this allows the primary to burn its $^{12} \rm C$ and $^{16}\rm O$ to higher amounts of $^{28}\rm Si - ^{40}\rm Ca$ because it is done at higher densities. The total nuclear energy of a SNe Ia can be approximated from its elemental composition as $ E_n \approx 1.55 (M_{\rm Ni}/M_{\astrosun}) + 1.18 (M_{\rm IME}/M_{\astrosun})$ \citep{branch92}. The kinetic energy of the ejecta is $E_{KE} \approx E_n-E_b$, where $E_b$ is the binding energy; and the three different models have the same binding energy because they have the same primary. Therefore, the extra nuclear energy of the higher density configurations results in a higher kinetic energy at homologous expansion.

\begin{figure}[th!]
\centering
\includegraphics[width=0.49 \textwidth]{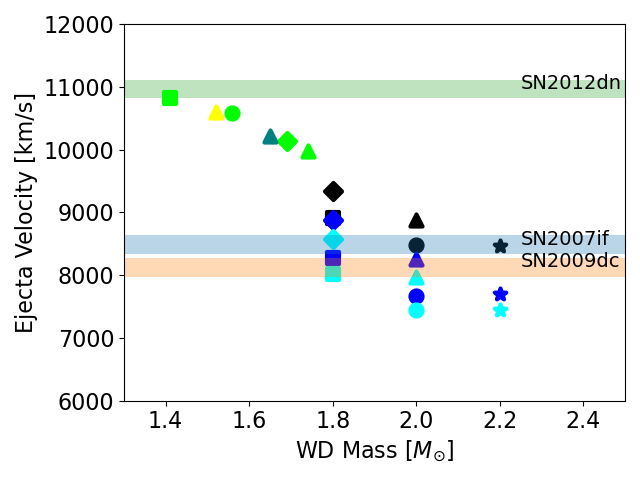}
\caption{Ejecta velocity as a function of total model mass for the merger models (blue and black points) and magnetized WD models (green), with symbols shown in Tables \ref{table:simulation_models_mergers} and \ref{table:simulation_models_magnetic}.We overplot the observed Si II velocities of the SC SNeIa SN2012dn, SN2007if, and SN2009dc at maximum light as shaded regions corresponding to an uncertainty of 300 km/s.}
\label{fig:ejecta_velocity}
\end{figure}

Figure \ref{fig:ejecta_velocity} shows the ejecta velocity at homologous expansion as a function of total model mass of all of the models run, $v_{KE}=\sqrt{2 E_{KE}/M_{\rm tot}}$, along with the Si II velocities of three SC SNeIa at maximum light overplotted as shaded regions \citep{ashall_carnegie_2021}. We show the ejecta velocity because it is strongly correlated with the linewidth velocities often inferred from observations. For the magnetized models, there is a positive correlation between the total model mass and both $E_n$ and $E_b$; therefore, the models are all roughly the same kinetic energy at homologous expansion and the higher mass models have a lower ejecta velocity. This trend follows similarly for merger models with the same primary. However, given the same \textit{total} mass for merger models, the velocity of the ejecta is higher for a higher primary mass, because the nuclear energy is higher. Overall, the merger models better match the low Si velocities inferred from the spectra of SC SNe Ia, which are often $<$ 10000 $\rm km s^{-1}$ at maximum light \citep{ashall_carnegie_2021}.

\subsection{Light Curves}

We now turn to the photometric properties of our models computed by SuperNu. Throughout the next section we primarily show magnetized model MAG5010 as our fiducial magnetized model and model MG101018 as our fiducial merger model, along with four other merger model variations. We show 1) model MG051018, to highlight the differences between two models differing only in primary mass, 2) model MG101022 to highlight the differences between two models differing only in secondary mass, and finally 3) models MG100518 and MG105018 to highlight the differences between two models differing between the three different density profiles. We discuss these models to demonstrate the properties of our model light curves; however, their properties are generally applicable to our other models as well.

To obtain the $K$-corrected observational data as well as pseudo bolometric lightcurves for all of the SC SNe Ia, we generated a template following the procedure in \cite{nugent_2002_templates}. This template was  constructed using the SNFactory data from SN~2012dn \citep{taubenberger_sn_2019}. \footnote{The Super-Chandrasekhar spectroscopic template is available \href{ https://c3.lbl.gov/nugent/nugent_templates.html}{here.}}

\begin{figure}[th!]
\centering
\includegraphics[width=0.49 \textwidth]{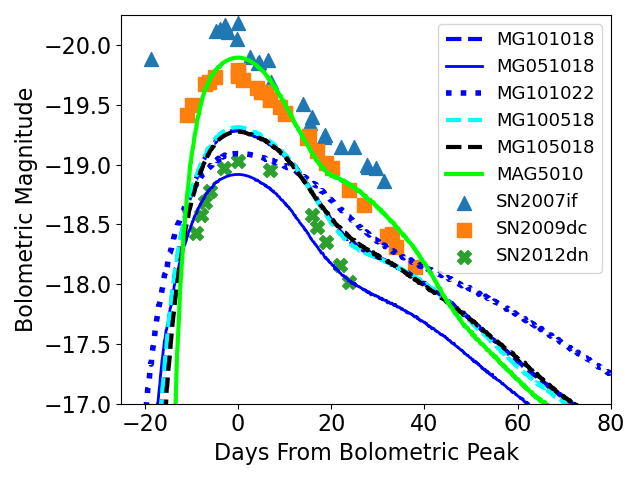}
\caption{Bolometric light curves for five of our merger models and one of our magnetized models. Overplotted are pseudo-bolometric light curves computed for the SC SNeIa SN2007if (blue triangles), SN2009dc (orange squares), and SN2012dn (green crosses).}
\label{fig:bolometric}
\end{figure}

Figure \ref{fig:bolometric} shows the bolometric light curves for these six models, along with pseudo-bolometric light curves for three SC SNe Ia (generated via the template as described above.) The magnetized models such as MAG5010 all have very similar light curves, with peak absolute bolometric magnitudes between -19.5 and -20 mag. Their light curves dim quickly after maximum light, declining by ~0.7-0.8 mag 15 days post maximum. The merger models show more variation. The peak brightness of their light curves is primarily a function of the amount of $^{56} \rm Ni$ synthesized in the explosion \citep{arnett_type_1982}, and therefore their primary mass. The MG10yyzz models such as MG101018 fall between -19 and -19.3 mag, and the MG05yyzz models such as MG051018 all fall between -18.8 and -19 mag. Their light curve shapes, in contrast, are primarily a function of the secondary mass; models with a higher mass secondary have a broader light curve. For example, models MG101018 and MG101022 shown in Figure \ref{fig:bolometric} have $\Delta m15$ values of 0.45 and 0.22 respectively. Finally, changing the density configuration of the secondary has no statistical difference on the light curve properties; Figure \ref{fig:bolometric} shows that all of the MG10yy18 models are visually identical. 

\begin{figure}[th!]
\centering
\includegraphics[width=0.49 \textwidth]{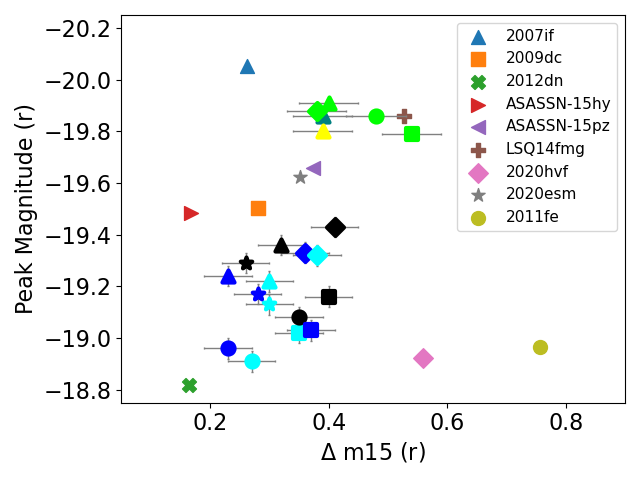}
\caption{The absolute $r$-band magnitude vs $\Delta m15 (r)$ for our merger models (blue and black points) and magnetized models (green points), with symbols shown in Tables \ref{table:simulation_models_mergers} and \ref{table:simulation_models_magnetic}. Error bars are taken as the median standard deviation. We overplot various SC SNe Ia and the normal SNe Ia SN 2011fe.}
\label{fig:phillips_relation}
\end{figure}

To compare our model light curves to observations, we plot the Phillips relation in the SDSS $r$-band in Figure \ref{fig:phillips_relation}, which shows the absolute $r$-band magnitude vs $\Delta m15 (r)$ for our models with SC SNe Ia overplotted. Though the Phillips relation is often plotted in the $B$-band, we chose to plot in the $r$-band because we found that the \textit{r}-band had the smallest offsets in the peak magnitude and light curve width from the bolometric quantities and the lowest spread in the offsets between the different models. The observational data are taken from \cite{scalzo_nearby_2010, taubenberger_high_2011, zhang_2016_SN2011fe, yamanaka_2016_2012dn, chen_2019_ASASSN_15pz, hsiao_2020_lsq14mg, lu_2021_ASASSN_15hy,  jiang_discovery_2021, dimitriadis_2022}.

Due to the large variation in the light curve properties of observed SC SNe Ia, most of our models fall reasonably within the observational results on the Phillips relation. The magnetized models have brighter peak luminosities comparable to the brighter SC SNe Ia such as SN 2007if and SN LSQ14fmg. Their $\Delta m15$ values are also relatively large but still fall within the observational data. The merger models have luminosities comparable to the dimmer SC SNe Ia such as SN 2012dn and SN 2009dc, and all of their $\Delta m15$ values fall within the observational data. Noticeably, all of our models fall far closer to SC SNe Ia observations than to `normal' SNe Ia such as SN 2011fe (yellow-green circle). 

\begin{figure*}[th!]
\centering
\includegraphics[width=0.98 \textwidth]{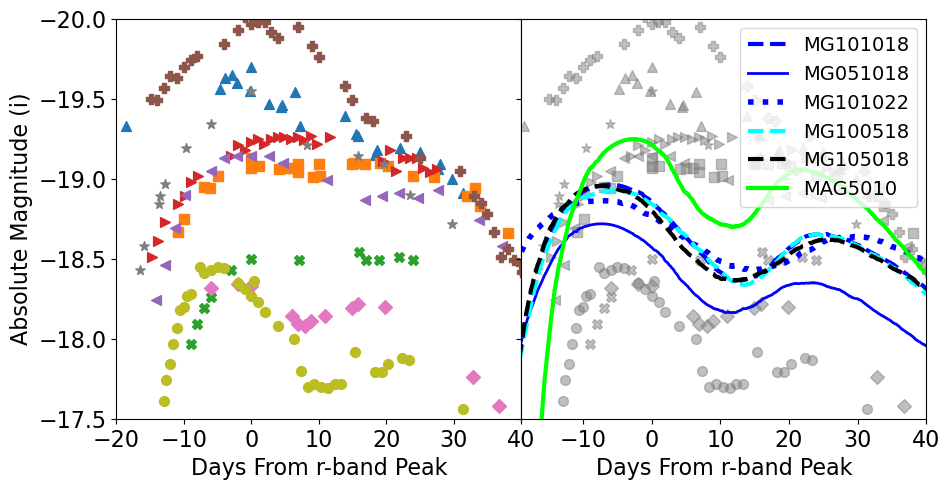}
\caption{Observational values of \textit{i}-band light curves (left) and five of our merger models and one of our magnetized models overplotted (right).}
\label{fig:iband}
\end{figure*}

We next turn to the individual photometric bands of SC SNe Ia, which have properties that distinguish them from normal SNe Ia. In general, SC SNe Ia do not have a prominent secondary maximum in the \textit{i}-band, which appears in the \textit{i}-band of normal SNe Ia due to the recombination of iron-group elements in the ejecta \citep{kasen_2006}. Figure \ref{fig:iband} shows the \textit{i}-band light curves of SC SNe Ia observations (left) and our models (right). Our models have \textit{i}-band luminosities comparable to SC SNe Ia observtaions, but the shape of the curves is similar to that of normal SNe Ia such as SN 2011fe (yellow-green circles). The magnetized models such as MAG5010 have especially prominent \textit{i}-band secondary maxima, rising nearly to the value at peak. The merger models have a less defined, but still noticeable secondary \textit{i}-band maximum. All of the models with the same secondary mass, such as the MGxxyy18 models shown in Figure \ref{fig:iband} have similarly shaped \textit{i}-band light curves. The model with a higher secondary mass, model MG101022, has a slightly flatter light curve, but still does not match observations. 

\begin{figure*}[th!]
\centering
\includegraphics[width=0.98 \textwidth]{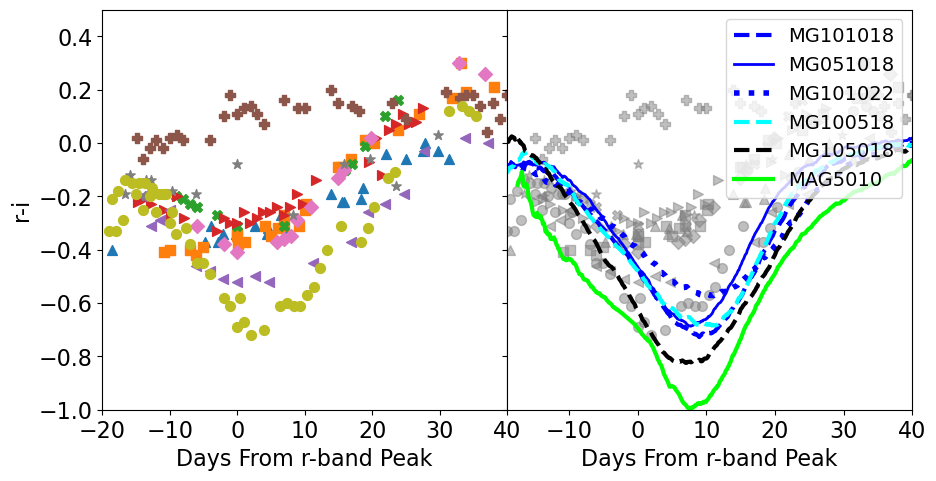}
\caption{Observational values of (\textit{r}-\textit{i}) light curves (left) and five of our merger models and one of our magnetized models overplotted (right).}
\label{fig:r_min_i}
\end{figure*}

Additionally, the (\textit{r}-\textit{i}) color curves of SC SNe Ia do not look like those of normal SNe Ia \citep{ashall_carnegie_2021}. Generally, they do not reach such large negative values. Figure \ref{fig:r_min_i} compares observations of SC SNe Ia (left) to our models (right). Our magnetized models such as MAG5010 are especially bright in the \textit{r}-band compared to the \textit{i}-band, so their $|$(\textit{r}-\textit{i})$|$ values are large. The merger models have smaller values of $|$(\textit{r}-\textit{i})$|$ which show minor variations between models varying secondary mass and density configuration. Figure \ref{fig:r_min_i} shows that models MG101022 and MG105018 have smaller and larger values of $|$(\textit{r}-\textit{i})$|$ respectively than model MG101018, indicating that models with a higher secondary mass and lower density configuration than their counterparts fall closer to observations. However, they still do not match obesrvational results for most SC SNe Ia. 

\subsection{Spectra}

\begin{figure*}[th!]
\centering
\includegraphics[width=0.98 \textwidth]{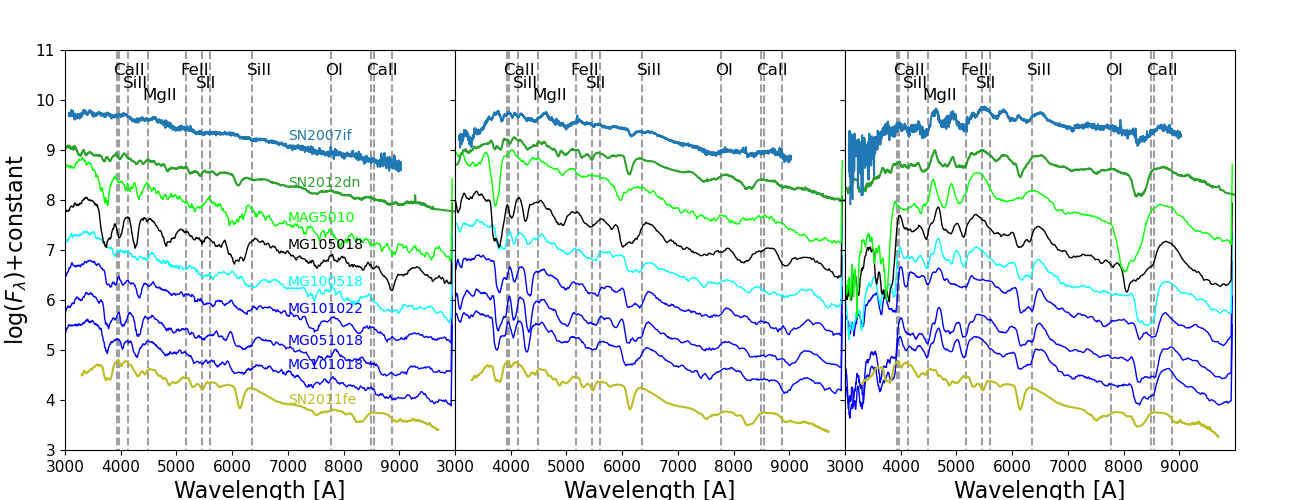}
\caption{Spectra at -10 days before maximum light (left panel), maximum light (middle panel) and +20 days after maximum light (right panel) for five of our merger models and one of our magnetized models computed with SuperNu, along with observations of the SC SNeIa SN 2007if and SN 2012dn and the normal SNe SN 2011fe at similar times. The observations are plotted at (-9, +5, +23) days from maximum light for SN 2007if, (-9, +0, +18) days from maximum light for SN 2012dn, and (-13, +7, +17) days for SN 2011fe. }
\label{fig:flux_comparison}
\end{figure*}

Finally, we turn to the synthetic spectra produced for our models with SuperNu. Figure \ref{fig:flux_comparison} shows the spectra from the models shown in previous figures at pre-maximum light, maximum light, and post-maximum light, along with the SC SNeIa SN2007if and SN2012dn  and the normal SNeIa SN2011fe at similar times. We note that all of our models have much deeper spectral lines than all of these observations, which is probably because they do not capture all of the physics of the interactions that occur in SC SNe Ia and non-LTE effects which may act to reduce the flux. However, we will still compare our model line depths to each other as it is a useful diagnostic tool for distinguishing SC SNe Ia. 

The left panel of Figure \ref{fig:flux_comparison} shows the spectra 10 days before maximum light. One of the easiest ways to distinguish SC SNe Ia from normal SNe Ia is by observing the spectra at this time, when they have comparatively weak, washed out features dominated by continuum and SiII absorption lines \citep{ashall_carnegie_2021}. Pre-maximum light, the SC SNe Ia SN 2012dn and SN 2007if are almost featureless compared to SN 2011fe. At this time, for our models the spectral features are primarily determined by the density configuration of the secondary, with the primary mass and secondary mass having little effect. As can be seen in Figure \ref{fig:flux_comparison}, model MG105018 has more prominent features than model MG101018, which has more features than model MG100518. In this respect, the low density configuration models match observations best. The density configuration also determines the line location; the velocity of the ejecta is higher for a higher secondary density (Figure \ref{fig:ejecta_velocity}), and so the spectra are more redshifted. Finally, we note that the magnetized models such as MAG5010 cannot be easily distinguished from the high and fiducial density merger models at this time.

At maximum light (middle panel of Figure \ref{fig:flux_comparison}), our merger models are more easily distinguished from our magnetized models. The SiII $\lambda 6355$ line is particularly prominent in our merger models at this time, where it appears saturated, often with a double minimum. This is caused by the two different regions of Si in the star, which can be seen in the right panel of Figure \ref{fig:mass_fractions}; that from the primary material and that from the secondary material, which are moving at different velocities. The magnetized models such as model MAG5010 have comparatively weak spectral features because they have lower amounts of intermediate mass elements than the merger models (Figure \ref{fig:mass_fractions_scatter}). Therefore, many of the transition lines which appear in the merger models such as the Si II $\lambda 4130$ and MgII $\lambda 4481$ do not appear in the spectra of model MAG5010 and our other magnetized models, and others such as the SiII $\lambda 6355$ are weaker.  Finally, post-maximum light (right panel of Figure \ref{fig:flux_comparison}), both classes of models show transition features from FeII, FeIII, and CoIII. The merger models continue to show the saturated features at the SiII $\lambda 6355$ line, which is also evident in the CaII $\lambda 8500$ features. 

\section{Discussion}
\label{section:discussion}

We have presented the results and analysis of a survey of 1D super-Chandrasekhar mass WD explosion models. To construct the models, we built density profiles for the WDs based off of two different observationally motivated predictions. The first class of models are WD merger models, constructed using a standard semi-relativistic polytrope model for the primary star and an additional layer of constant density material on top for the secondary. The second class of models are magnetized WD models, constructed using the model outlined in \cite{bhattacharya_2022_magnetic}. We modeled the thermonuclear explosion of the models using the hydrodynamics code Castro, and produced model light curves and spectra using the radiation transport code SuperNu. The following characteristics summarize their properties:

1. The amount of radioactive $^{56} \rm Ni$ produced in the explosions, which is the primary determinant of the SNe light curve brightness, is very different for the two classes of models even with relatively similar masses. The magnetized models burn almost entirely to $^{56} \rm Ni$ (left panel of Figure \ref{fig:mass_fractions}. The merger models do not burn any of their secondary to $^{56} \rm Ni$ because the density is too low (right panel of Figure \ref{fig:mass_fractions}); and consequently, the luminosities of their light curves do not depend on the properties of the secondary at all.

2. Both classes of models fall within the range of observations of SC SNe Ia on the Phillips relation (Figure \ref{fig:phillips_relation}). The WD merger models fall closer to lower luminosity SC SNe Ia such as SN 2012dn and SN ASASSN-15hy, while the magnetized models fall closer to brighter SC SNe Ia such as SN 2007if and SN LSQ14fmg. 

3. The ejecta velocities of the merger models are comparable to those obtained from observations of SC SNe Ia spectroscopic linewidths, while the magnetized models have higher ejecta velocities comparable to normal SNe Ia observations (Figure \ref{fig:ejecta_velocity}). 

4. Neither class of models reproduces the photometric properties that are the defining characteristics of SC SNe Ia, namely the lack of a double peak in the $i$ band (Figure \ref{fig:iband}) and an $(r-i)$ color curve that is less negative than normal SNe Ia (Figure \ref{fig:r_min_i}). However, the merger models come closer, with slight variation based on the model properties.  
 
Given these remarks, we end with the conclusion that SC SNe Ia probably do not originate from the explosion of a magnetized WD without a companion. Despite thoroughly exploring the parameter space of magnetized WDs proposed by \cite{bhattacharya_2022_magnetic}, we were not able to construct a magnetized model of a single WD that reproduces the spectroscopic and photometric properties of SC SNe Ia observations; instead, those were more akin to very bright normal SNe Ia. The merger models come closer, and the reasons they do not match observations completely may be primarily due to drawbacks in our radiation transport, such as not modeling all reactions sufficiently. Additionally, SuperNu does not have non-LTE capabilities, which may change the observables if included. Non-LTE physics has never been explored for SC SNe Ia; however, as shown by \cite{shen_2021}, including non-LTE in SNe Ia radiative transfer calculations can produce significant differences in the light curves and spectra. Given these considerations, a WD merger event alone may be sufficient to explain the lower luminosity SC SNe Ia such as SN 2012dn.

However, the reader may wonder what the origin is of the very luminous SC SNe Ia such as SN 2007if and SN LSQ14fmg. After all, we thoroughly explored the parameter space of realistic WD merger models that satisfy the conditions necessary for a merger event and subsequent explosion, namely: 1) both WDs are CO WDs, 2) $M_{SD} < M_{P}$, and 3) $M_{\rm tot} > M_{\rm ch}$. We speculate these very bright measurements may be caused by a non-spherical explosion that increases the luminosity preferentially in one direction, which is not captured in our 1D models. Polarization measurements of SC SNe Ia such as those presented in \cite{cikota_2019} for normal SNe Ia will help determine whether this is the case. Another possibility is that these events are due to a WD merger event in which the primary (and possibly secondary) star is magnetized, and has $M_{P} > M_{\rm ch}$. This may be sufficient to produce enough $^{56} \rm Ni$ to match the brightest luminosity observations, while also preserving the spectroscopic and photometric properties of the merger models we studied. This avenue will be explored in future work. 

\acknowledgements

Funding for this research came from the Director, Office of Science, Office of High Energy Physics of the U.S. Department of Energy under Contract no. DE-AC02-05CH1123. The National Energy Research Scientific Computing Center, a DOE Advanced Scientific Computing User Facility under the same contract, provided staff, computational resources, and data storage for this project. This material is based upon work supported by the U.S. Department of Energy, Office of Science, Office of Advanced Scientific Computing Research, Department of Energy Computational Science Graduate Fellowship under Award Number DE-SC0021110. We would also like to thank Kate Maguire, Georgios Dimitriadis, and Maxime Deckers for several useful conversations on more recent observational work in this field. 


\bibliography{supernova_references}

\end{document}